\documentclass[preprint,review,12pt,sort&compress]{elsarticle}
\usepackage[margin=1.18 in]{geometry}
\usepackage{multirow}
\usepackage[usenames, dvipsnames]{color}
\definecolor{UW}{RGB}{64, 38, 96}

\usepackage{amssymb}

\usepackage{float}

\journal{Composites Science and Technology}

\begin{document}

\begin{titlepage}

\clearpage\thispagestyle{empty}



\noindent

\hrulefill

\begin{figure}[h!]

\centering

\includegraphics[width=1.5 in]{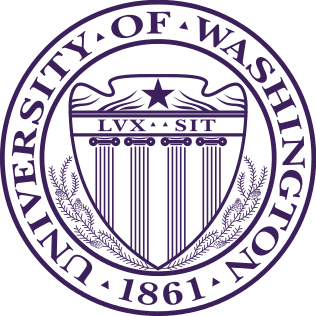}

\end{figure}


\begin{center}

{\color{UW}{

{\bf A\&A Program in Structures} \\ [0.1in]

William E. Boeing Department of Aeronautics and Astronautics \\ [0.1in]

University of Washington \\ [0.1in]

Seattle, Washington 98195, USA

}

}

\end{center} 

\hrulefill \\ \vskip 2mm

\vskip 0.5in

\begin{center}

{\large {\bf A Study on the Fracturing Behavior of Thermoset Polymer Nanocomposites}}\\[0.5in]

{\large {\sc Yao Qiao, Cory Hage Mefford, Marco Salviato}}\\[0.75in]

{\sf \bf INTERNAL REPORT No. 17-07/02E}\\[0.75in]

\end{center}

\noindent {\footnotesize {{\em Submitted to Composites Science and Technology \hfill July 2017} }}

\end{titlepage}

\newpage

\begin{frontmatter}


\cortext[cor1]{Corresponding Author, \ead{salviato@aa.washington.edu}}

\title{A Study on the Fracturing Behavior of Thermoset Polymer Nanocomposites}


\author[address]{Yao Qiao}
\author[address]{Cory Hage Mefford}
\author[address]{Marco Salviato\corref{cor1}}

\address[address]{William E. Boeing Department of Aeronautics and Astronautics, University of Washington, Seattle, Washington 98195, USA}

\begin{abstract}
\linespread{1}\selectfont

This work proposes an investigation on the fracturing behavior of polymer nanocomposites. Towards this end, the study leverages on the analysis of a large bulk of fracture tests from the literature with the goal of critically investigating the effects of the nonlinear Fracture Process Zone (FPZ). 

It is shown that for most of the fracture tests, the effects of the nonlinear FPZ are not negligible, leading to significant deviations from Linear Elastic Fracture Mechanics (LEFM). As the data indicate, this aspect needs to be taken into serious consideration since the use of LEFM to estimate mode I fracture energy, which is common practice in the literature, can lead to an error as high as $157$\% depending on the specimen size and nanofiller content.

\end{abstract}


\end{frontmatter}


\section{Introduction}
\label{intro}
The outstanding advances in polymer nanocomposites in recent years have paved the way for their broad use in engineering. Potential applications of these materials include microelectronics \cite{Rogers}, energy storage \cite{Yoo} and harvesting \cite{chih}, soft robotics \cite{zhaoxuanhe}, and bioengineering \cite{Joong}. One of the reasons of this success is that, along with remarkable enhancements of physical properties such as e.g. electric and thermal conductivity \cite{Ramirez,Balandin}, nanomodification offers significant improvements of stiffness \cite{jiang2013_1}, strength \cite{konnola2015_1} and toughness \cite{zappalorto2013_1,zappalorto2013_2,CoryandYao}. These aspects make it an excellent technology to enhance the mechanical behavior of polymers \cite{zhang2008_1,johnsen2007_1,carolan2016_1,zamanian2013_1,dittanet2012_1,liu2011_1,WaJin13,ChaSei13,ChaSa14,kim2008_1,naous2006_1,wetzel2006_1,vaziri2011_1} or to improve the weak matrix-dominated properties of fiber composites \cite{pathak}.

While a large bulk of data on the mechanical properties of polymer nanocomposites is available already, an aspect often overlooked is the effect on the fracturing behavior of the region close to the crack tip featuring most of energy dissipation, the \textit{Fracture Process Zone} (FPZ). This is an important aspect since, due to the complex mesostructure characterizing nanocomposites, the size of the non-linear FPZ occurring in the presence of a large stress-free crack is usually not negligible \cite{Baz84,Baz90,bazant1996_1,bazant1998_1,salviato2016_1} leading to a significant deviation from the typical brittle behavior of thermoset polymers. This phenomenon cannot be captured by classical Linear Elastic Fracture Mechanics (LEFM) which inherently assumes the size of the FPZ to be negligible compared to the structure size. To seize the effects of a finite, non-negligible FPZ, the introduction of a characteristic (finite) length scale related to the fracture energy and the strength of the material is necessary \cite{Baz84,Baz90,bazant1998_1,bazant1996_1,salviato2016_1}.

This work proposes an investigation on the fracturing behavior of thermoset polymer nanocomposites with the goal of critically investigating the effects of the nonlinear Fracture Process Zone (FPZ). By employing Size Effect Law (SEL), a formulation endowed with a characteristic length inherently related to the FPZ size, and assuming a linear cohesive behavior \cite{cusatis2009_1}, a large bulk of literature data is analyzed. It is shown that for most of the fracture tests, the nonlinear behavior of the FPZ is not negligible, leading to significant deviations from LEFM. As the data indicate, this aspect needs to be taken into serious consideration since the use of LEFM to estimate mode I fracture energy can lead to an error as high as $157$\% depending on the specimen size and nanofiller content.

\section{Quasi-brittle Fracture of Nanocomposites} 
\label{sec:literature}
In nanocomposites, the size of the non-linear Fracture Process Zone (FPZ) occurring in the presence of a large stress-free crack is generally not negligible. The stress field along the FPZ is nonuniform and decreases with crack opening, due to a number of damage mechanisms such as e.g. discontinuous cracking, micro-crack deflection, plastic yielding of nanovoids, shear banding and micro-crack pinning \cite{zhang2008_1,salviato2011_2, salviato2013_1,salviato2013_2,zappalorto2013_1,zappalorto2013_2,Schulte_14, quaresimin2014_1, Quaresimin_16,CoryandYao}. As a consequence, the fracturing behavior and, most importantly, the energetic size effect associated with the given structural geometry, cannot be described by means of classical Linear Elastic Fracture Mechanics (LEFM) which assumes the effects of the FPZ to be negligible. To capture the effects of a finite, non-negligible FPZ, the introduction of a characteristic (finite) length scale related to the fracture energy and the strength of the material is necessary \cite{Baz84,Baz90,bazant1998_1,bazant1996_1,salviato2016_1}. This is done in the following sections.

\subsection{Size effect law for nanocomposites}
\label{sec:literatureapplication}
\begin{figure} [H]
\center
\includegraphics[scale=0.8]{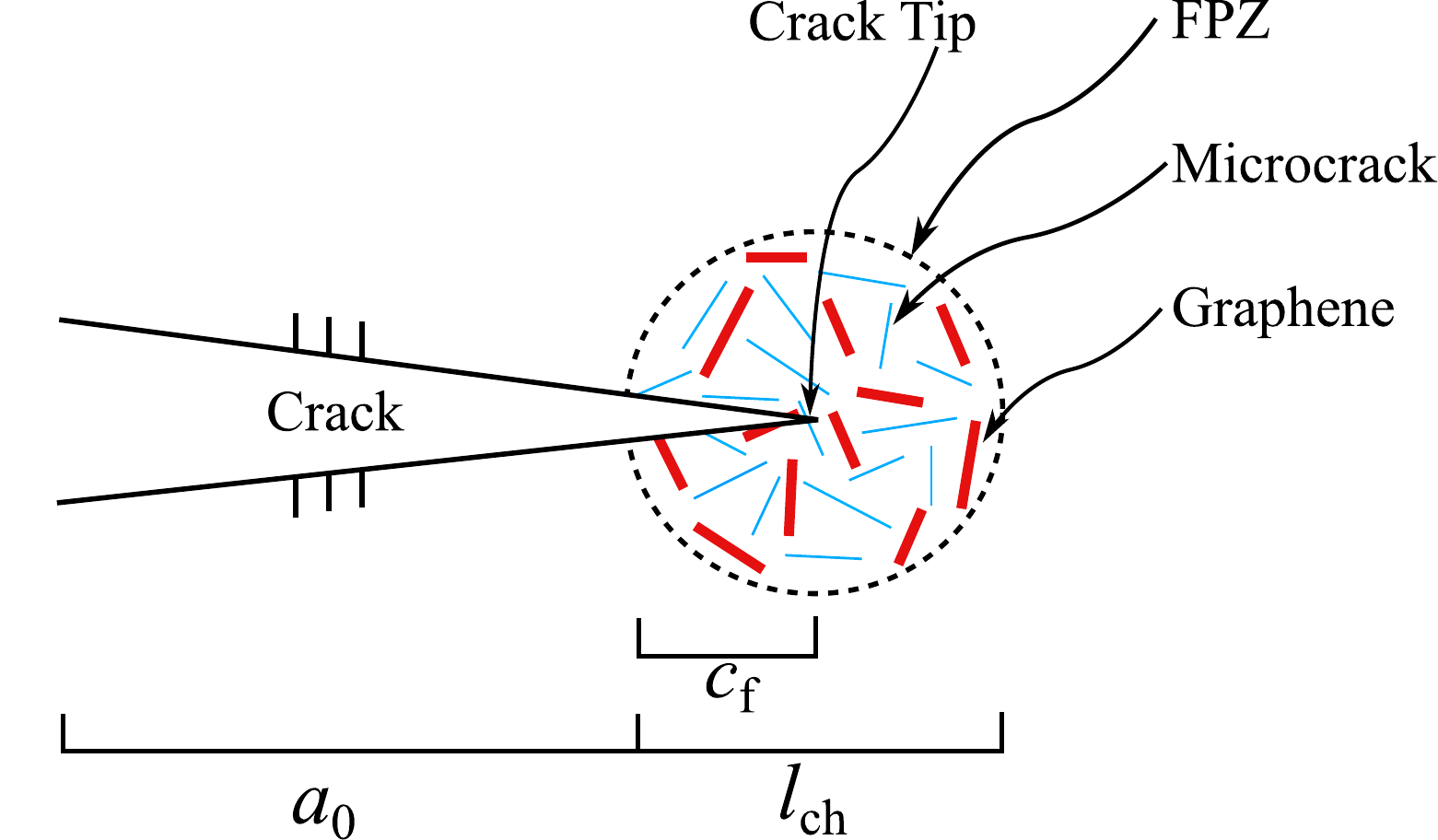}
\caption{Fracture Process Zone (FPZ) for thermoset polymer nanocomposites.}
\label{fig:FPZexample}
\end{figure}

\begin{figure} [H]
\center
\includegraphics[scale=0.5]{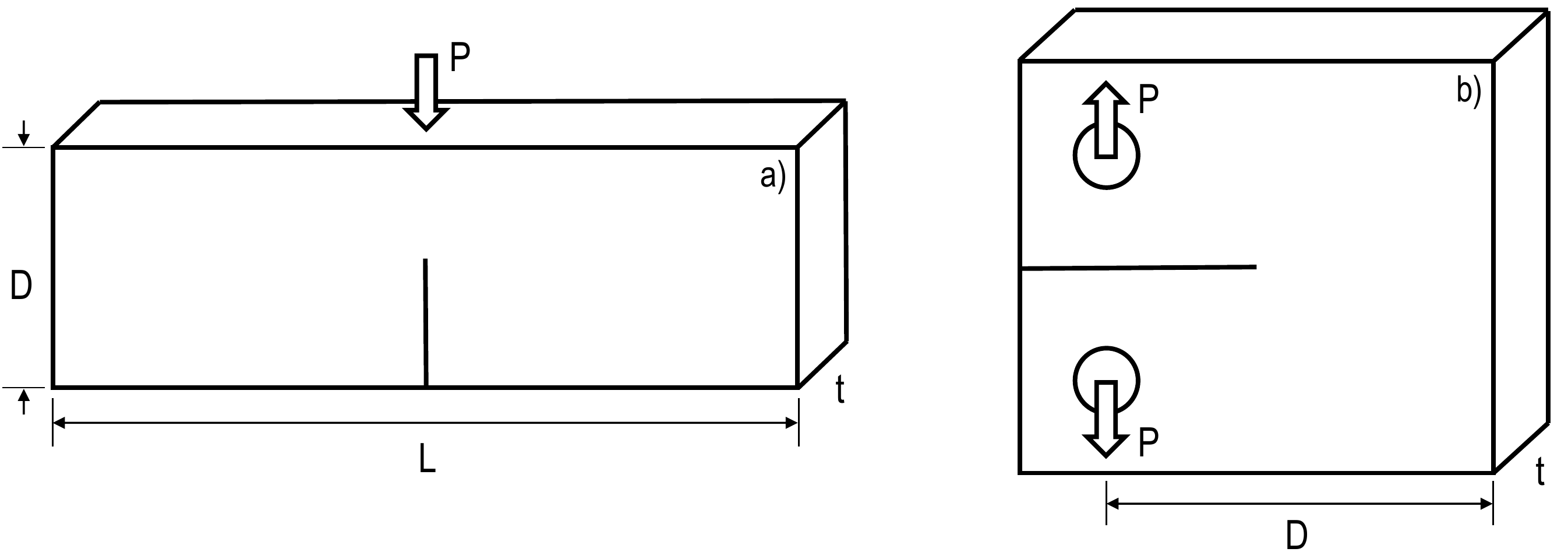}
\caption{Schematic representation of the SENB and CT specimens considered in this work.}
\label{fig:newgeometries}
\end{figure}

The fracture process in nanocomposites can be analyzed leveraging on an equivalent linear elastic fracture mechanics approach to account for the presence of a FPZ of finite size as shown in Fig. \ref{fig:FPZexample}. To this end, an effective crack length $a=a_0+c_f$ with $a_0=$ initial crack length and $c_f=$ effective FPZ length is considered. Following LEFM, the energy release rate can be written as follows:
\begin{equation}
G\left(\alpha\right)=\frac{\sigma_N^2D}{E^*}g(\alpha)
\label{eq:Gf}
\end{equation}
where $\alpha=a/D=$ normalized effective crack length, $E^*= E$ for plane stress and $E^*= E/\left(1-\nu^2\right)$ for plane strain, $g\left(\alpha\right)=$ dimensionless energy release rate and, $D$ is represented in Fig. \ref{fig:newgeometries} for Single Edge Notch Bending (SENB) and Compact Tension (CT) specimens respectively. $\sigma_N$ represents the nominal stress defined as e.g. $\sigma_N=3PL/2tD^2$ for SENB specimens or $\sigma_N=P/tD$ for CT specimens where, following Fig. \ref{fig:newgeometries}, $P$ is the applied load, $t$ is the thickness and $L$ is the span between the two supports for a SENB specimen as defined in ASTM D5045-99 \cite{ASTM_SENB}.

At incipient crack onset, the energy release rate ought to be equal to the fracture energy of the material. Accordingly, the failure condition can now be written as:
\begin{equation}
G\left(\alpha_0+c_f/D\right)=\frac{\sigma_{Nc}^2D}{E^*}g\left(\alpha_0+c_f/D\right)=G_f
\label{failure}
\end{equation}
where $G_f$ is the mode I fracture energy of the material and $c_f$ is the effective FPZ length, both assumed to be material properties. It should be remarked that this equation characterizes the peak load conditions if $g'(\alpha)>0$, i.e. only if the structure has positive geometry \cite{bazant1998_1}.

By approximating $g\left(\alpha\right)$ with its Taylor series expansion at $\alpha_0$ and retaining only up to the linear term of the expansion, one obtains:
\begin{equation}
G_f=\frac{\sigma_{Nc}^2D}{E^*} \left[g(\alpha_0)+\frac{c_f}{D}g'(\alpha_0)\right]
\label{Taylor}
\end{equation}
which can be rearranged as follows \cite{bazant1998_1}:
\begin{equation}
\sigma_{Nc}=\sqrt{\frac{E^*G_f}{Dg(\alpha_0)+c_fg'(\alpha_0)}}
\label{eq:Sel}
\end{equation}
where $g'\left(\alpha_0\right)=\mbox{d}g\left(\alpha_0\right)/\mbox{d}\alpha$.

This equation, known as Ba\v zant's Size Effect Law (SEL) \cite{Baz84,Baz90,bazant1998_1,salviato2016_1}, relates the nominal strength to mode I fracture energy, a characteristic size of the structure, $D$, and to a characteristic length of the material, $c_f$, and it can be rewritten in the following form:
\begin{equation}
\sigma_{Nc}=\frac{\sigma_{0}}{\sqrt{1+D/D_0}}
\label{eq:sigmaNc2}
\end{equation}
with $\sigma_0=\sqrt{E^*G_f/c_fg'(\alpha_0)}$ and $D_0=c_fg'(\alpha_0)/g(\alpha_0)=$ constant, depending on both FPZ size and specimen geometry. Contrary to classical LEFM, Eq. (\ref{eq:sigmaNc2}) is endowed with a characteristic length scale $D_0$. This is key to describe the transition from ductile to brittle behavior with increasing structure size.

\subsection{Calculation of $g\left(\alpha\right)$ and $g'\left(\alpha\right)$}
\label{sec:literatureforgeneralnanocomposites}

\subsubsection{Single Edge Notch Bending (SENB) specimens}
\label{sec:literatureSENB}
The calculation of $g(\alpha)$ and $g'(\alpha)$ for SENB specimens can be done according to the procedure described in \cite{CoryandYao}. This leads to the following polynomial expressions:
\begin{equation}
g(\alpha)=1155.4\alpha^5-1896.7\alpha^4+1238.2\alpha^3-383.04\alpha^2+58.55\alpha-3.0796
\end{equation}
\begin{equation}
g'(\alpha)=18909\alpha^5-31733\alpha^4+20788\alpha^3-6461.5\alpha^2+955.06\alpha-50.88
\end{equation}

\subsubsection{Compact Tension (CT) specimens}
\label{sec:literatureCT}
In the case of CT specimens, the values for $g(\alpha)$ and $g'(\alpha)$ can be determined leveraging on the equations provided by ASTM D5045-99 \cite{ASTM_SENB}. Following the standard, the mode I Stress Intensity Factor (SIF), $K_I$, can be written as:
\begin{equation}
K_I=\frac{P}{t\sqrt{D}}f(\alpha)
\label{eq:K1cCT} \\
\end{equation}
where $\alpha=a/D$ and $D$ is the distance between the center of hole to the end of the specimen as defined in ASTM D5045-99 \cite{ASTM_SENB} (see Fig. \ref{fig:newgeometries}b). The nominal stress $\sigma_N$ for CT specimens can be defined as:
\begin{equation}
\sigma_N=\frac{P}{tD}
\label{eq:stressnomCT}
\end{equation} 
The mode I Stress Intensity Factor can be rewritten as follows by combining Eq. (\ref{eq:K1cCT}) and Eq. (\ref{eq:stressnomCT}):
\begin{equation}
K_I=\sqrt{D}\sigma_Nf(\alpha)
\label{eq:K1cCT2}
\end{equation}
By considering the relationship between energy release rate and stress intensity factor for a plane strain condition, the mode I energy release rate results into the following expression:
\begin{equation}
G_I=\frac{D\sigma^2_{N}}{E^*}g(\alpha) 
\label{eq:gCT}
\end{equation} 
where $g(\alpha)=f^2(\alpha)(1-\upsilon^2)$, and $f(\alpha)$ is a dimensionless function accounting for geometrical effects and the finiteness of the structure (see e.g. \cite{ASTM_SENB}). Once $g(\alpha)$ is derived, the expression of $g'(\alpha)$ can be obtained by differentiation leading to the following polynomial expressions for $g(\alpha)$ and $g'(\alpha)$ respectively:
\begin{equation}
g(\alpha)=33325\alpha^5-52330\alpha^4+32016\alpha^3-9019.1\alpha^2+1230.1\alpha-51.944
\end{equation}
\begin{equation}
g'(\alpha)=555868\alpha^5-895197\alpha^4+554047\alpha^3-159153\alpha^2+21035\alpha-917.3
\end{equation}

\section{Fracture behavior of thermoset nanocomposites: analysis and discussion}
In the following sections, several data on the fracturing behavior of nanocomposites are critically analyzed employing the expressions derived in Section \ref{sec:literature}. First, some recent tests on geometrically-scaled SENB specimens made of a thermoset polymer reinforced by graphene are reviewed to investigate how the FPZ affects the failure behavior. Then, leveraging on SEL and assuming a linear cohesive behavior, a large bulk of data from the literature originally elaborated by LEFM is re-analyzed to include the effects of the FPZ.

\subsection{Fracture Scaling of Graphene Nanocomposites} 
\label{sec:graphene}
To investigate the effects of the non-linear FPZ, it is useful to review some recent evidences on the scaling of the fracturing behavior. To this end, the fracture tests on geometrically-scaled SENB specimens reported by Mefford \textit{et al}. \cite{CoryandYao}, who studied a thermoset polymer reinforced by graphene nanoplatelets, are analyzed and discussed. 

\begin{figure} [H]
\center
\includegraphics[scale=0.5]{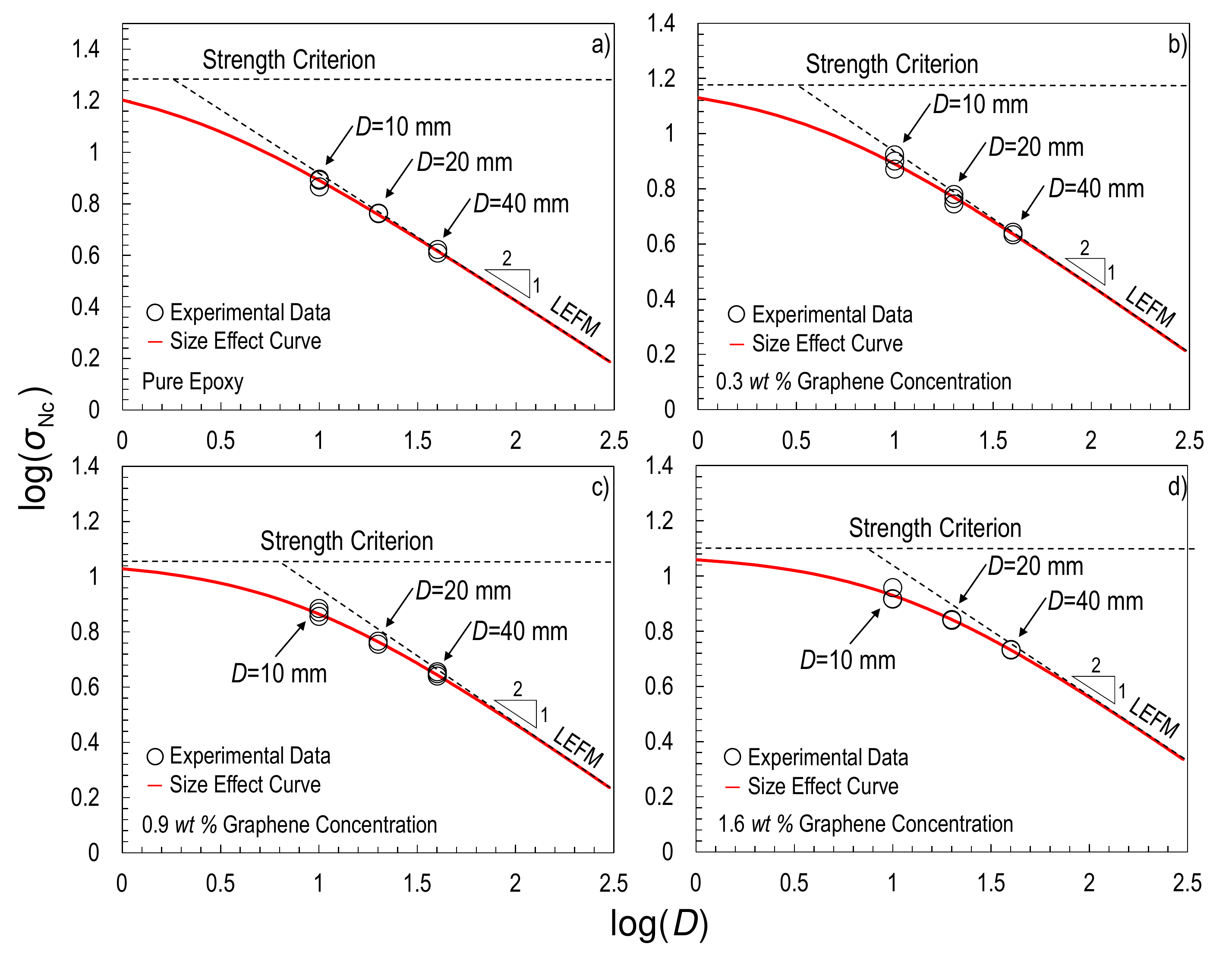}   
\caption{Size effect curves for different graphene concentrations obtained by testing geometrically-scaled SENB specimens of different sizes \cite{CoryandYao}.}
\label{fig:sizeeffectcurves}
\end{figure}

Figures \ref{fig:sizeeffectcurves}a-d show the experimental structural strength $\sigma_{Nc}$ and the fitting by SEL plotted as a function of the structure size $D$ in double logarithmic scale.   In such a graph, the structural scaling predicted by LEFM is represented by a line of slope $-1/2$ whereas the case of no scaling, as predicted by stress-based failure criteria, is represented by a horizontal line. The intersection between the LEFM asymptote, typical of brittle behavior, and the plastic asymptote, typical of ductile behavior, corresponds to $D=D_0$, called the \emph{transitional size} \cite{bazant1998_1}. The figure reports the size effect tests for various graphene contents, from the case of a pristine polymer to $wt\%=1.6$. 

As can be noted from Figure \ref{fig:sizeeffectcurves}a, the experimental data related to the pure epoxy system all lie very close to the LEFM asymptote showing that, for the range of sizes investigated (or larger sizes), linear elastic fracture mechanics provides a very accurate description of fracture scaling. This shows that, for the pure epoxy and sufficiently large specimens, the FPZ size has a negligible effect and LEFM can be applied, as suggested by ASTM D5045-99 \cite{ASTM_SENB}. However, this is not the case for graphene nanocomposites which, as Figures \ref{fig:sizeeffectcurves}b-d show, are characterized by a significant deviation from LEFM, the deviation being more pronounced for smaller sizes and higher graphene concentrations. In particular, the figures show a transition of the experimental data from stress-driven failure, characterized by the horizontal asymptote, to energy driven fracture characterized by the $-1/2$ asymptote. This phenomenon can be ascribed to the increased size of the FPZ compared to the structure size which makes the non-linear effects caused by micro-damage in front of the crack tip not negligible. For sufficiently small specimens, the FPZ affects the structural behavior and causes a significant deviation from the scaling predicted by LEFM with a much milder effect of the size on the structural strength. On the other hand, for increasing sizes, the effects of the FPZ become less and less significant thus leading to a stronger size effect closely captured by LEFM. Further, comparing the size effect plots of nanocomposites with different graphene concentrations, it can be noted a gradual shift towards the ductile region thus showing that not only the addition of graphene leads to a higher fracture toughness but also to a gradually more ductile structural behavior for a given size. 

As the experimental data show, LEFM does not always provide an accurate method to extrapolate the structural strength of larger structures from lab tests on small-scale specimens, especially if the size of the specimens belonged to the transitional zone. In fact, the use of LEFM in such cases may lead to a significant underestimation of structural strength, thus hindering the full exploitation of graphene nanocomposite fracture properties. This is a severe limitation in several engineering applications such as e.g. aerospace or aeronautics for which structural performance optimization is of utmost importance. On the other hand, LEFM always overestimates significantly the strength when used to predict the structural performance at smaller length-scales. This is a serious issue for the design of e.g. graphene-based MEMS and small electronic components or nanomodified carbon fiber composites in which the inter-fiber distance occupied by the resin is only a few micrometers and it is comparable to the FPZ size. In such cases, SEL or other material models characterized by a characteristic length scale ought to be used.

\subsection{Effects of a finite FPZ on the calculation of Mode I fracture energy}
\label{sec:literatureforgeneralnanocomposites}
Notwithstanding the importance of understanding the scaling of the fracturing behavior, the tests conducted by Mefford \textit{et al}. \cite{CoryandYao} represent, to the best of the authors' knowledge, the only comprehensive investigation on the size effect in nanocomposites available to date. All the fracture tests reported in the literature were conducted on one size and analyzed by means of LEFM. Considering the remarkable effects of the nonlinear FPZ on the fracturing behavior documented in the foregoing section, it is interesting to critically re-analyze the fracture tests available in the literature by means of SEL. This formulation is endowed with a characteristic length related to the FPZ size and, different from LEFM, it has been shown to accurately capture the transition from brittle to quasi-ductile behavior of nanocomposites.

\subsubsection{Application of SEL to thermoset polymer nanocomposites}
\label{sec:literatureapplication}
To understand if the quasi-brittle behavior reported in previous tests \cite{CoryandYao} is a salient feature of graphene nanocomposites only or if it characterizes other nanocomposites, a large bulk of literature data were re-analyzed by SEL using Eq.(\ref{Taylor}) in order to study the effects of the FPZ. In this analysis, in the absence of data on the effective FPZ length, $c_f$, in the literature, it is assumed that $c_f=0.44 l_{ch}$ which, according to Cusatis \emph{et al.} \cite{cusatis2009_1}, corresponds to the assumption of a linear cohesive law. In this expression, $l_{ch}=E^*G_f/f_t^2$ is Irwin's characteristic length which depends on Young's modulus $E^*$, the mode I fracture energy $G_f$ and the ultimate strength of the material $f_t$. Substituting this expression into Eq. (\ref{Taylor}) and rearranging, one gets the following expression which relates the fracture energy calculated according to SEL to the fracture energy calculated by LEFM:

\begin{equation}
G_{f,SEL}=\frac{G_{f,LEFM}}{1-\frac{0.44 E^*g'(\alpha_0)G_{f,LEFM}}{Df_t^2 g(\alpha_0)}}
\label{eq:Gf_SEL2}
\end{equation}
In this equation,  $G_{f,LEFM}=\sigma_{Nc}^2 Dg(\alpha_0)/E^*$ represents the fracture energy which can be estimated by analyzig the fracture tests by LEFM (note that this expression lacks of a characteristic length scale). 

It can be observed from Eq.(\ref{eq:Gf_SEL2}) that, once $g(\alpha)$ and $g^{'}(\alpha)$ are calculated, the fracture energy corrected for the effects of the FPZ can be calculated by knowing three key parameters: (1) the fracture energy estimated by LEFM, (2) the Young's modulus of the specimen, and (3) the ultimate strength of the specimens at different nanofiller concentrations. For cases in which those parameters were not provided by the authors, the ultimate strength, Young's modulus, and Poisson's ratio were reasonably assumed to be $50$ MPa, $3000$ MPa, and $0.35$ respectively. 

\subsubsection{Mode I fracture energy of thermoset polymer nanocomposites}
\begin{figure}[H]
\center
\includegraphics[scale=0.5]{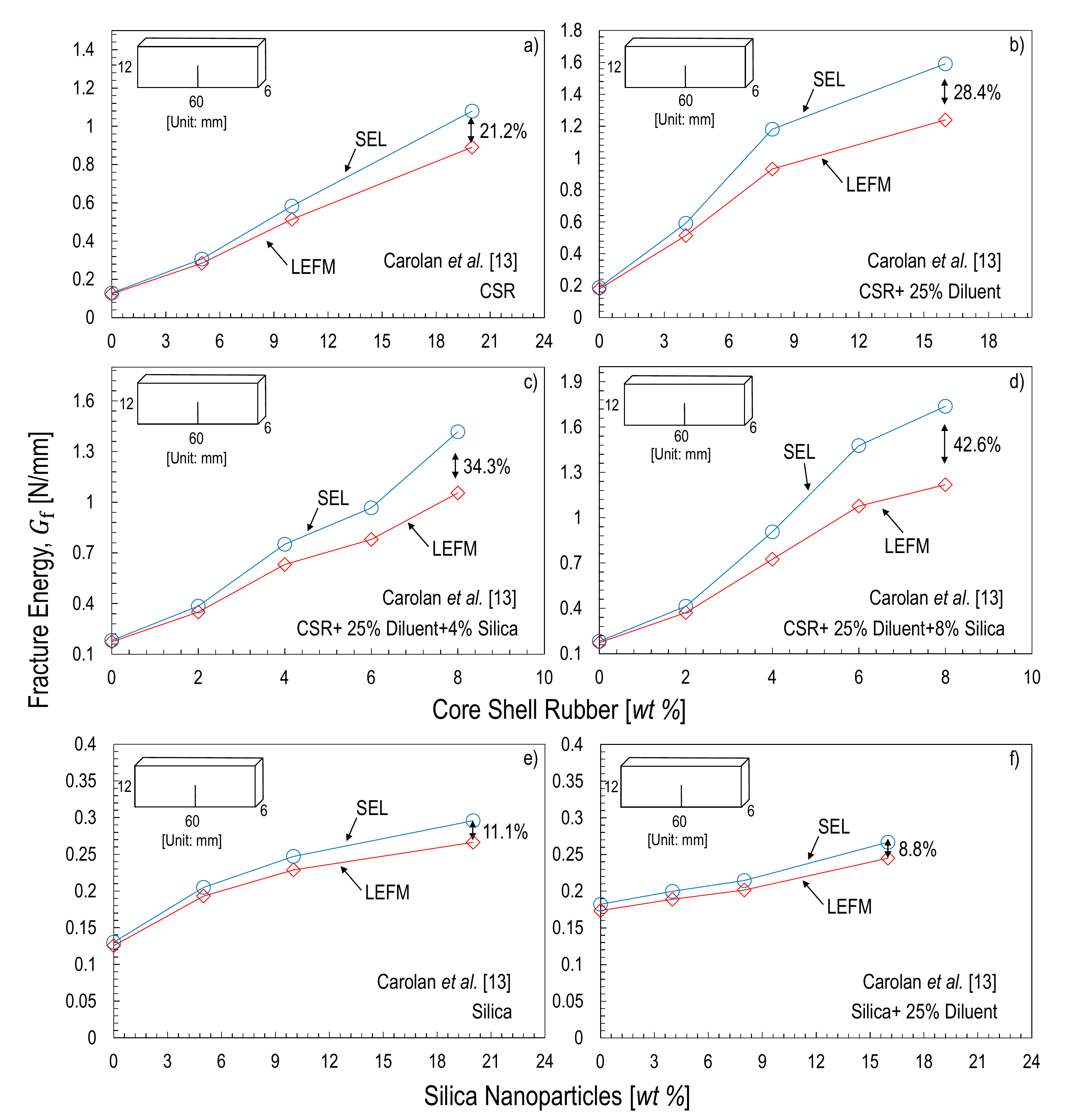} 
\caption{Mode I fracture energy estimated by Linear Elastic Fracture Mechanics (LEFM) and Size Effect Law (SEL), Eq. (\ref{eq:Gf_SEL2}). The latter formulation accounts for the finite size of the nonlinear Fracture Process Zone (FPZ) in thermoset nanocomposites. Data re-analyzed from \cite{carolan2016_1}.}
\label{fig:carolan}
\end{figure}

\begin{figure}[H]
\center
\includegraphics[scale=0.5]{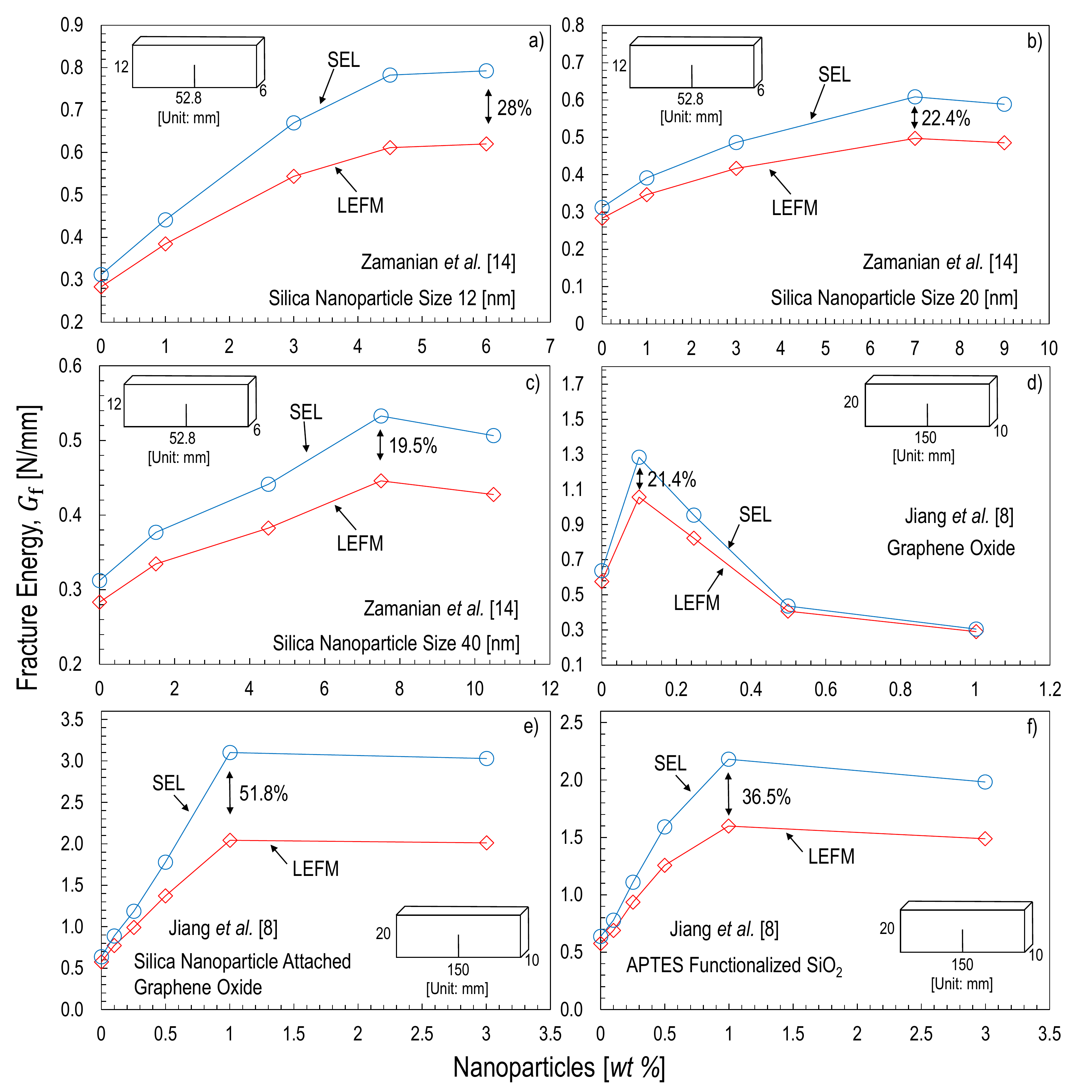}
\caption{Mode I fracture energy estimated by Linear Elastic Fracture Mechanics (LEFM) and Size Effect Law (SEL), Eq. (\ref{eq:Gf_SEL2}). The latter formulation accounts for the finite size of the nonlinear Fracture Process Zone (FPZ) in thermoset nanocomposites. Data re-analyzed from \cite{zamanian2013_1} and \cite{jiang2013_1}.}
\label{fig:zamanian&jiang}
\end{figure}

\begin{figure}[H]
\center
\includegraphics[scale=0.5]{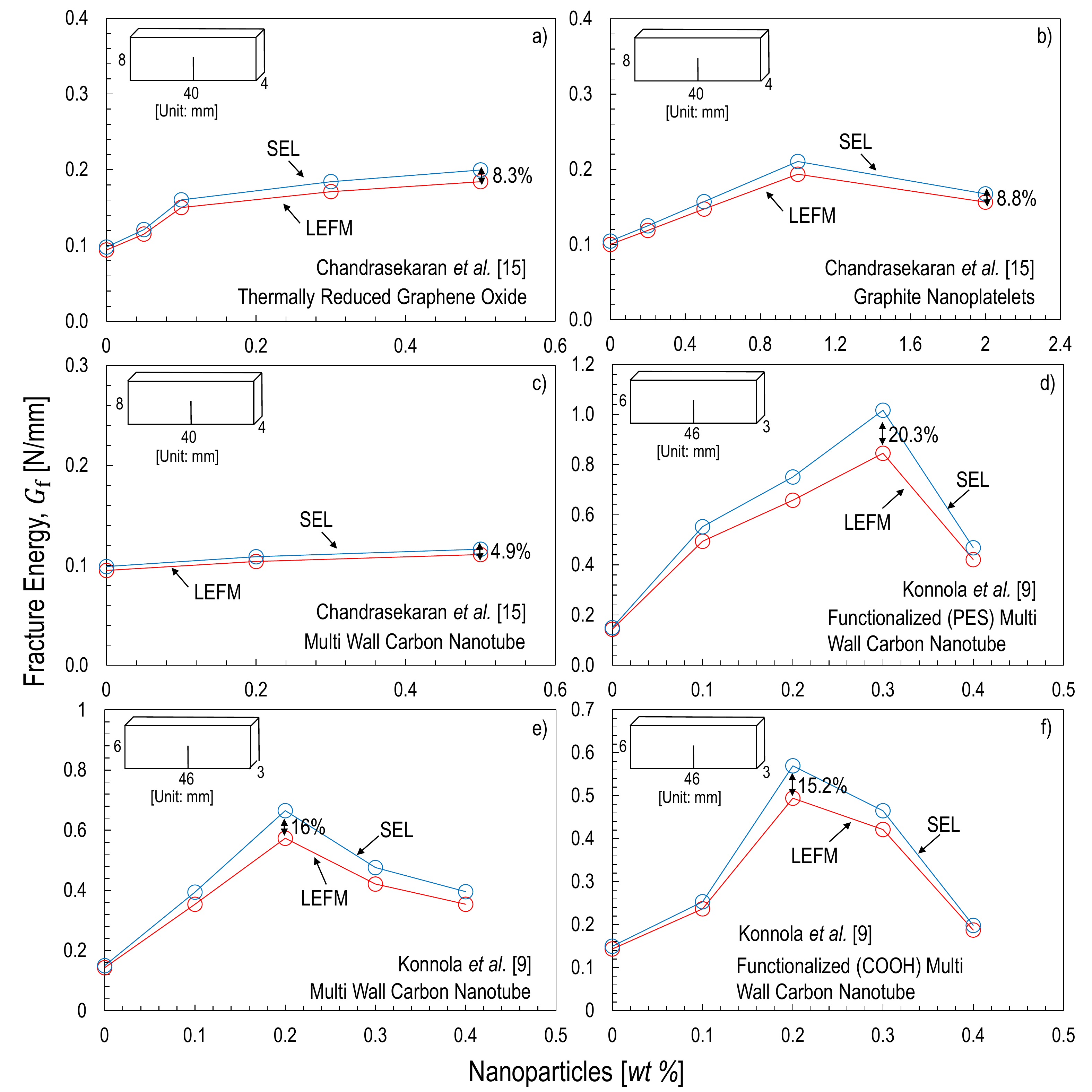}
\caption{Mode I fracture energy estimated by Linear Elastic Fracture Mechanics (LEFM) and Size Effect Law (SEL), Eq. (\ref{eq:Gf_SEL2}). The latter formulation accounts for the finite size of the nonlinear Fracture Process Zone (FPZ) in thermoset nanocomposites. Data re-analyzed from \cite{ChaSa14} and \cite{konnola2015_1}.}
\label{fig:chandra&konnola}
\end{figure}

\begin{figure}[H]
\center
\includegraphics[scale=0.5]{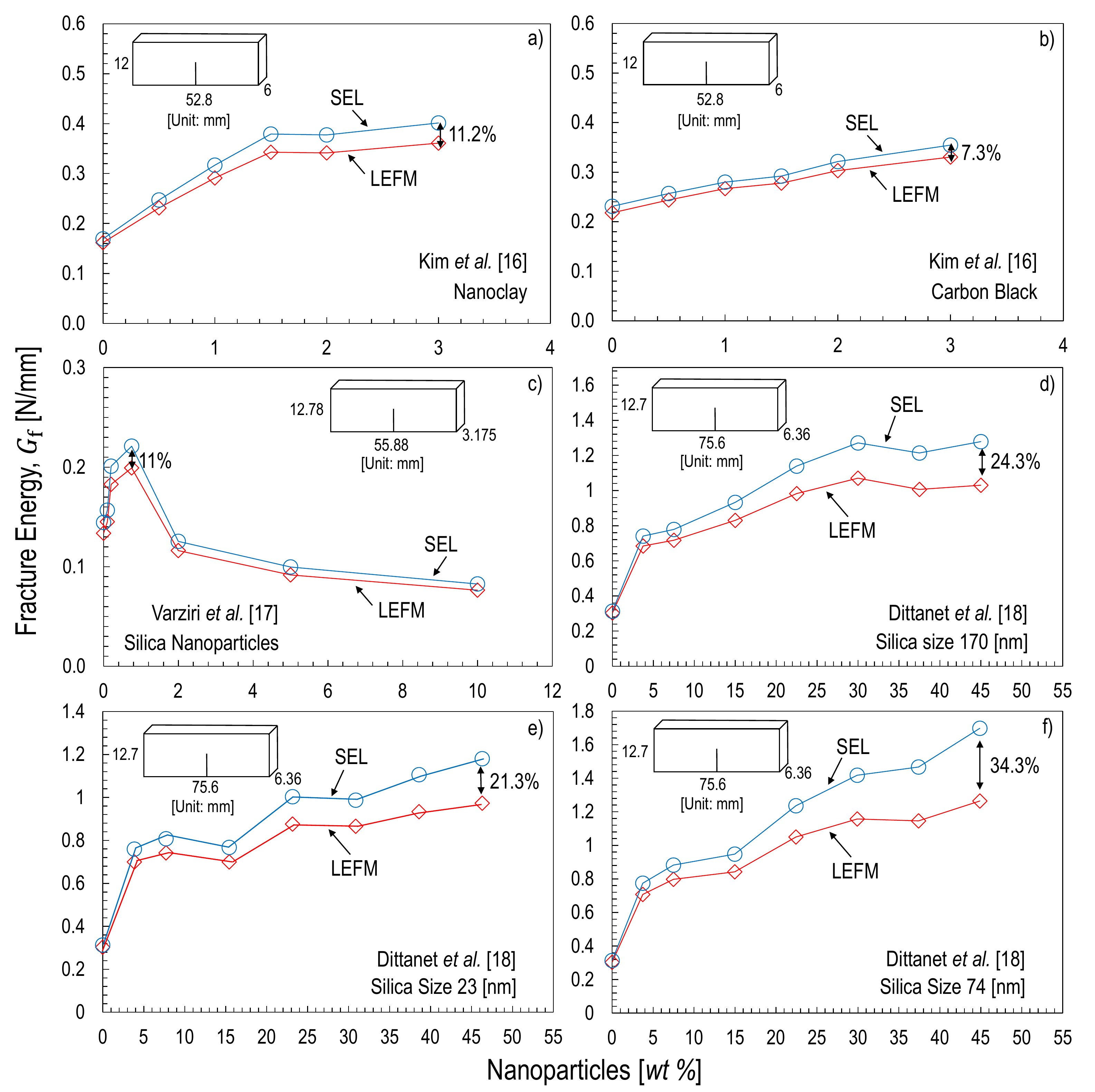}
\caption{Mode I fracture energy estimated by Linear Elastic Fracture Mechanics (LEFM) and Size Effect Law (SEL), Eq. (\ref{eq:Gf_SEL2}). The latter formulation accounts for the finite size of the nonlinear Fracture Process Zone (FPZ) in thermoset nanocomposites. Data re-analyzed from \cite{kim2008_1}, \cite{vaziri2011_1} and \cite{dittanet2012_1}.}
\label{fig:dittanet&vaziri&kim}
\end{figure}

\begin{figure}[H]
\center
\includegraphics[scale=0.5]{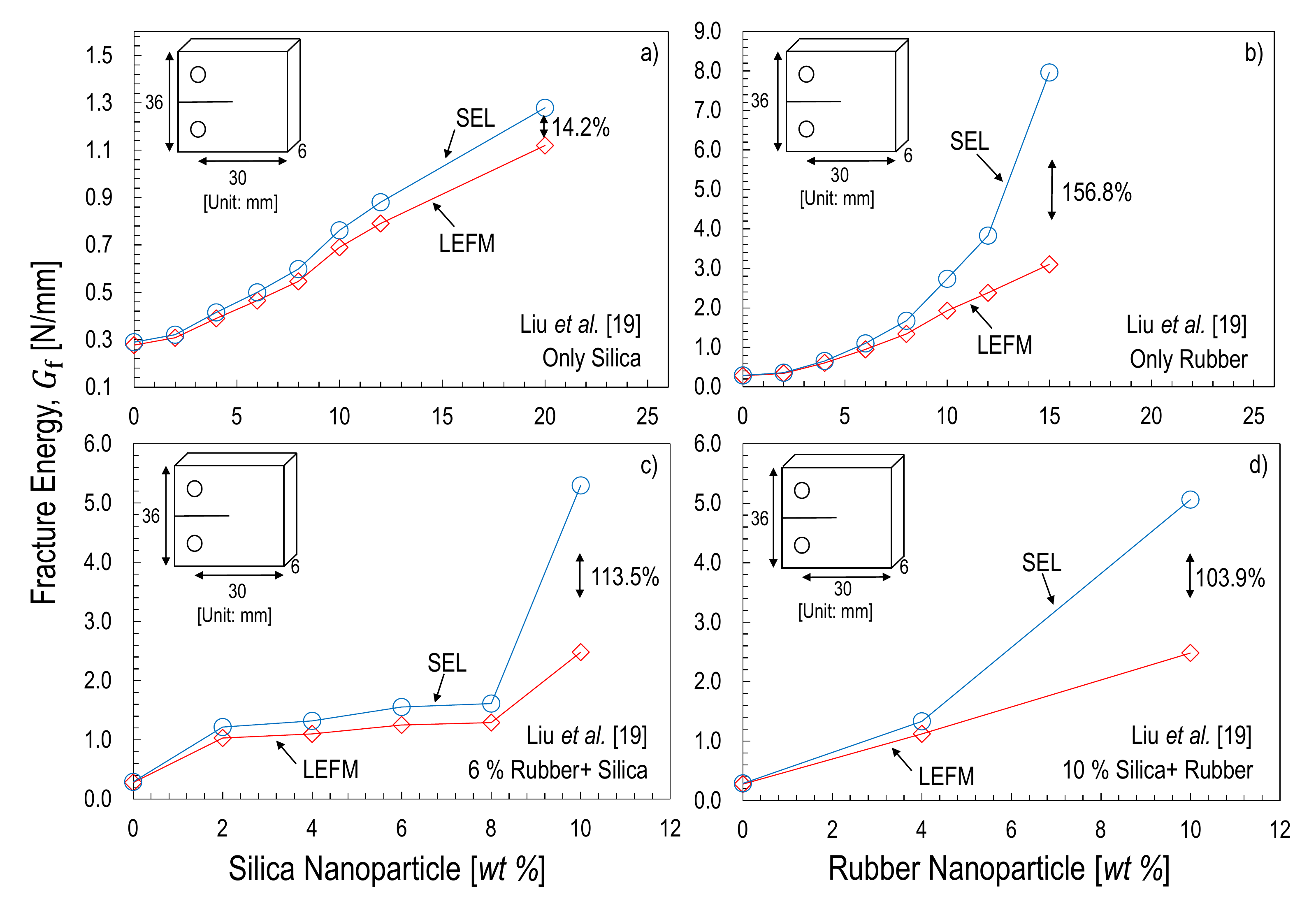}
\caption{Mode I fracture energy estimated by Linear Elastic Fracture Mechanics (LEFM) and Size Effect Law (SEL), Eq. (\ref{eq:Gf_SEL2}). The latter formulation accounts for the finite size of the nonlinear Fracture Process Zone (FPZ) in thermoset nanocomposites. Data re-analyzed from \cite{liu2011_1}.}
\label{fig:liu}
\end{figure}

Several types of nanofillers were investigated in this re-analysis including carbon-based nano-fillers (such as carbon black, graphene oxide, graphene nanoplatelets, and multi-wall carbon nanotubes), rubber and silica nanoparticles, and nanoclay. The fracture energy estimated from LEFM compared to the calculation through SEL, Eq. (\ref{eq:Gf_SEL2}), for nanomodified SENB and CT specimens are plotted in Figures \ref{fig:carolan}-\ref{fig:liu} along with the highest difference.

Figure \ref{fig:carolan} shows data elaborated from Carolan \emph{et al.} \cite{carolan2016_1} who conducted fracture tests on SENB specimens nano-modified by six different combinations of nanofillers. As can be noted, while for the pristine polymer the difference between LEFM and SEL is negligible, this is not the case for the nanomodified polymers, the difference increasing with increasing nanofiller content. The difference varies based on the type of nanofiller used, with the greatest value being 42.6\% for the addition of 8 wt\% core shell rubber mixed with 25\% diluent and 8\% silica. This remarkable discrepancy confirms that for the SENB specimens tested in \cite{carolan2016_1} the nonlinear behavior of the FPZ was not negligible, leading to a more ductile behavior compared to the pristine polymer. 

Similar conclusions can be drawn based on Figures \ref{fig:zamanian&jiang}a-f which report the analysis of fracture tests conducted by Zamanian \emph{et al.} \cite{zamanian2013_1} and Jiang \emph{et al.} \cite{jiang2013_1} on polymers reinforced by silica nanoparticles and silica nanoparticle$+$graphene oxide respectively. For the data in \cite{zamanian2013_1}, the greatest percent difference of fracture energy between LEFM and SEL decreased as the size of silica nanoparticle increased, with the greatest difference being 28\% for the addition of 6 wt\% 12 nm silica nanoparticles. For all the systems investigated, the maximum deviation from LEFM was for the largest amount of nanofiller, confirming that nanomodification lead to larger FPZ sizes and more pronounced ductility. On the other hand, the data by Jiang \emph{et al.} \cite{jiang2013_1} exhibit an even larger effect of the FPZ with the greatest difference in fracture energy between LEFM and SEL reaching up to 51.8\% for silica nanoparticle attached to graphene oxide. 

A milder effect of the FPZ can be inferred from the data by Chandrasekaran \emph{et al.} \cite{ChaSa14} who investigated three types of carbon-based nano-fillers (Figure \ref{fig:chandra&konnola}): (1) thermally reduced graphene oxide; (2) graphene nanoplatelets; and (3) multi-wall carbon nanotubes. In these cases, the difference between SEL and LEFM ranges from 4.9\% to 8.8\%, the lowest difference among all the data analyzed in this study. For these systems, the specimen size compared to the size of the nonlinear FPZ was large enough to justify the use of LEFM which provided accurate and objective results. On the other hand, a more significant effect of the FPZ can be inferred from the data reported by Konnola \emph{et al.} \cite{konnola2015_1} who studied three different types of functionalized and nonfunctionalized nano-fillers. In this case, the greatest difference in fracture energy ranges between 15.2\% to 20.3\%.

SENB specimens nano-modified by nanoclay and carbon black respectively were tested by Kim \emph{et al.} \cite{kim2008_1}. As Figure \ref{fig:dittanet&vaziri&kim} shows, in this case, the specimen size was enough to justify the use of LEFM as confirmed by the low difference with SEL (11.2\% for nanoclay and 7.3\% for carbon black). Similar conclusions can be drawn on the silica nanoparticles investigated by Vaziri \emph{et al.} \cite{vaziri2011_1}. However, for the three different sizes of silica nanoparticles investigated by Dittanet \emph{et al.} \cite{dittanet2012_1}, a significant difference between LEFM and SEL was observed, confirming that these specimens tested belonged to the transition zone between ductile and brittle behavior where the effects of the nonlinear FPZ cannot be neglected.

Figure \ref{fig:liu} shows a re-analysis of the data reported by Liu \emph{et al.} \cite{liu2011_1} who tested CT specimens nano-modified by four different combinations of silica nanoparticle and rubber. As can be noted, in this case, the FPZ indeed affects the fracturing behavior significantly. Adopting LEFM, which assumes the size of the FPZ to be negligible, for the estimation of $G_f$ from the fracture tests would lead to an underestimation of up to 156.8\% for the case of polymer reinforced by 15 wt\% rubber only. This tremendous difference, the largest found in the present study, gives a tangible idea on the importance of accounting for the nonlinear damage phenomena occurring in nanocomposites which can lead to a significant deviation from the typical brittle behavior of thermoset polymers.

\section{Conclusions}
Leveraging on a large bulk of literature data, this paper investigated the effects of the Fracture Process Zone (FPZ) on the fracturing behavior of thermoset polymer nanocomposites, an aspect of utmost importance for structural design but so far overlooked. Based on the results obtained in this study, the following conclusions can be elaborated:

1. The fracture scaling of pure thermoset polymers is captured accurately by Linear Elastic Fracture Mechanics (LEFM). However, this is not the case for nanocomposites which exhibit a more complicated scaling. The double logarithmic plots of the nominal strength as a function of the characteristic size of geometrically-scaled SENB specimens \cite{CoryandYao} showed that the fracturing behavior evolves from ductile to brittle with increasing sizes. For sufficiently large specimens, the data tend to the classical $-1/2$ asymptote predicted by LEFM. However, for smaller sizes, a significant deviation from LEFM was reported with data exhibiting a milder scaling, a behavior associated to a more pronounced ductility. This trend was more and more pronounced for increasing nanofiller contents;

2. Following Ba\v{z}ant \cite{Baz84,Baz90,bazant1998_1}, an Equivalent Fracture Mechanics approach can be used to introduce a characteristic length, $c_f$, into the formulation. This length is related to the FPZ size and it is considered a material property as well as $G_f$. The resulting scaling equation, known as Ba\v{z}ant's Size Effect Law (SEL), depends not only on $G_f$ but also on the FPZ size. An excellent agreement with experimental data is shown, with SEL capturing the transition from quasi-ductile to brittle behavior with increasing sizes. 

3. By employing Size Effect Law and assuming a linear cohesive behavior \cite{cusatis2009_1}, a large bulk of literature data on the mode I fracture energy of thermoset nanocomposites was critically re-analyzed. It is shown that for most of the fracture tests in the literature, the effects of the nonlinear FPZ are not negligible, leading to significant deviations from LEFM. As the data indicate, this aspect needs to be taken into serious consideration since the use of LEFM to estimate mode I fracture energy can lead to an error as high as $156$\% depending on the specimen size and nanofiller content.

4. The deviation from LEFM reported in the re-analyzed results is related to the size of the Fracture Process Zone (FPZ) for increasing contents of nanofiller. In the pristine polymer the damage/fracture zone close to the crack tip, characterized by significant non-linearity due to subcritical damaging, was generally very small compared to the specimen sizes investigated. This was in agreement with the inherent assumption of LEFM of negligible non-linear effects during the fracturing process. However, the addition of nano-fillers results in larger and larger FPZs. For sufficiently small specimens, the size of the highly non-linear FPZ was not negligible compared to the specimen characteristic size thus highly affecting the fracturing behavior, this resulting into a significant deviation from LEFM;
 
5. The foregoing evidences show that particular care should be devoted to the fracture characterization of nanocomposites. LEFM, which inherently assumes the FPZ to correspond to a mathematical point, can only be used to estimate mode I fracture energy when the specimen size is large enough compared to the FPZ. For small specimens, for which the energy dissipated in the nonlinear FPZ contributes significantly on the overall energy in the structure, a formulation endowed with a characteristic size related to the FPZ ought to be used. Alternately, as was shown in \cite{CoryandYao}, size effect testing on geometrically scaled specimens represents a simple and effective approach to provide objective estimates of the fracture energy.




\section*{Acknowledgments}
Marco Salviato acknowledges the financial support from the Haythornthwaite Foundation through the ASME Haythornthwaite Young Investigator Award and from the University of Washington Royalty Research Fund. This work was also partially supported by the William E. Boeing Department of Aeronautics and Astronautics as well as the College of Engineering at the University of Washington through Salviato's start up package.


\newpage
\section*{Figures}

\newpage

\newpage

\newpage

\newpage

\newpage

\newpage


\end{document}